# A Proposal for the Establishment of Review Boards – a flexible approach to the selection of academic knowledge

Bruce Edmonds

*Centre for Policy Modelling,*
*Manchester Metropolitan University*
http://www.cpm.mmu.ac.uk/~bruce

## 1. Introduction

An academic researcher today is faced with many of the same basic problems as those encountered be researchers 50 years ago. These include: finding the relevant work done by other academics, being able to disseminate their work to other academics and gaining the recognition due for it. The nature of these difficulties has changed – the difficulties of access and cost used to dominate academic publishing whereas now academics have access to so much information that they are swamped. New techniques are constantly being developed in order to enable academics to efficiently select the information they want and filter out the rest.

The traditional remedy is the peer-reviewed journal. A relatively small number of trusted academics select what they judge to be worthwhile for the rest to read. The system has its limitations, for example if a reader does not have the same selection criteria as the referees and editor, the available journals may refuse papers this reader would have wanted to read. This restriction was acceptable because of the expense and time it saved. Now, with the advent of cheap distribution via the internet a *new* trade-off between time and expense and the flexibility of the selection process is possible.

This paper explores one such possible process – one where the role of mark-up and archiving is separated from that of review. The idea is that authors publish their papers on their own web pages or in a public paper archive, a board of reviewers judge that paper on a number of different criteria. The results of the reviews are stored in such a way as to enable readers to use these judgements to find the papers they want using search engines on the web.

The following section (section 2) looks at some recent innovations that use the internet to aid the distribution of knowledge. These innovations pave the way for the main proposal, which is described in section 3. Section 5 relates the proposal to some different pictures of knowledge development. Section 4 then examines some of the possible consequences of the proposal. Section 6 lists some of the practical steps needed in order to implement the proposal and I conclude in section 7.

## 2. Mechanisms of academic knowledge distribution using the internet

Paper-based journals have significant costs associated with them, including: writing the papers; reviewing them; organisation of the review process; mark-up; printing; distribution; archiving; indexing; search by the readers and the reading. Some of these are of the essence of the process and should not be eliminated. However, the advent of available computational power and the internet means that some of this cost structure is different for electronic means of publication. I consider four briefly, noting their advantages and shortcomings.

## 2.1. Web-journals

Traditionally organised web-journals eliminate the need for printing (at least by the publisher), significantly reduce the distribution and archiving costs, and facilitate the organisation of the review process, the indexing and the reader's search. What costs there are can be almost totally subsumed into the academic's time and resources (internet access, printing facilities, word-processing). This means that the only major unsubsumed costs left are the organisation of review, and the mark-up of papers. Since a lot of the mark-up can be demanded of the writer using word-processing facilities, the remaining unsubsumed costs are sufficiently low that web-journals can be set up with no explicit commercial structure at all. Also the flexible nature of electronic representation of knowledge can be utilised to enable new forms of knowledge presentation and interaction between academics, for example in peer-commentary.

However web-journals still have some the rôles of paper journals, in that these journals still own the papers they publish. They mark them up to ensure they are well presented and so that the style of papers is that of the journal, they archive the papers on their web-site and (typically) own the copyright of their papers. This ownership has subtle costs in that it makes the structure of knowledge less flexible – for example it means that a single paper can only be presented to one audience.

Another disadvantage is in the closed nature of the review process. There is a private dialogue between the reviewers and editors of the journal and the author. Frequently this dialogue concerns not only questions of fact and presentation but also the content of the paper, even when the issues concerned are controversial and far from settled. This not only deprives the public readership of a part in this discussion but also denies them the information that can be gained from it (e.g. the detail of the reviewers judgements). As a result the readers can only search journals on the content of what is published – in a real sense the *only* explicit judgemental information given to the reader is that it is worthy of being published.

## 2.2. Search engines

The explosion in the amount of information available on the web means that is increasingly impractical to merely 'surf' around, if you are in need of a particular type of information. Internet search engines allow one to considerably increase the chance of finding information if it is out there, and one's search target is quite specific. However, it does not allow one to filter on quality, standard of presentation or the like. Thus they are very difficult to use if one wants information on a topic upon which there are many available pages. Further more this is a problem that is unlikely to be able to be fixed by purely technical solutions, since the competition for readers in popular areas means that the web authors will constantly adapt to exploit the qualities of these web engines whatever they are.

## 2.3. Web review services

A number of services have sprung up to provide judgement upon web-sites. Examples include Magellen which provides a rating service and Encyclopedia Brittanica which simply reviews and selects what it considers to be quality web-sites. Subscribers who use these can be assured of a certain type and level of quality when they choose among the selected sites. However, their generic nature means that they will not be able to cover individual academic papers in sufficent depth to be useful to academics.

## 2.4. Paper archives

Following Paul Ginsparg's ftp archive for physics papers [8], a number of public academic paper archives have been set up. Academics are free to up-load any papers they have written to the site which then archives them. The advantage of this over just having the papers on individual's faculty or personal home pages is that they can be stored in a consistent way so that they can be effectively searched using a search engine. The reader may feel they can rely on the continued existence and URL of these papers, so that they can reliably refer to them there. Finally 'alert' services can be added so that readers can be notified by e-mail when new papers arrive in their chosen categories. These archives do not give any guarantee of quality, although only academics tend to post papers to them and some idea of quality can be guessed at by the author, the title, the institution of the author etc. However one can not restrict the search to quality papers only, one can only access them on their explicit content.

## 2.5. On-line Indexing Services

Academics have compiled indexes to aid other academics for hundreds of years. Such indexes are more convenient to access when available on-line. A few of these services (especially specialist services) include links to on-line papers so that the results of searches can be immediately accessed. Inclusion in such an index does give a substantial, if indirect, guarantee of quality (they usually rely on the judgements of the journals they think are reliable). However the inclusion of papers still gives little indication about levels of quality (although some are approaching this problem through citation information and statistics).

## 2.6. Review Services associated with Public Archives

There are at least two peer-review services that are proposed or being set up in conjunction with public paper archives. The first is the American Physical Society's (APS) plan to review papers on the Ginsparg's physics pre-print archive. The idea is that the APS will organize the peer-review of papers on the arvchive that are submitted to them. The final version of accepted papers will then be put on to the archive along with a APS certificate. The second is a proposal by Chistopher Bauma and Thomas Krichel as part of their application for funding of their Economics Distributed Electronic Library (EDEL) project [2], which is designed to strengthen and extend the RePIC service [15].

Both of these aim to add a qulaity mark to the papers on the public archive and, in this way, to allow readers to make a basic judgement as to whether papers are worthwhile reading. However, the process is almost as costly as that associated with web-journals and only provides limited judgemental information on the papers reviewed (essentially a single bit). For this reason it does not provide readers with the ability to tailor their own criteria for papers in a flexible manner.

# 3. A Proposal for the Establishment of Electronic Review Boards

The basic idea is that review board will be set up which will review on-line papers (whether on home pages or in archives). The reviews will be done by a board of reviewers, who would aim to cover a particular area of knowledge (and maybe only a particular aspect of such an area). The reviews would consist of judgements of the papers in a number of different ways. A paper could be given a grade from one to five on presentation, relevance, soundness of argument, originality, importance of questions considered, importance of results etc. as well as short comments about the paper for public consumption. These reviews would be submitted electronically and automatically collated into a set of publicly accessible records. These records would include information about the content (title, author, abstract, keywords etc.) as well as the compiled qualitative judgements (number or reviewers, average grades, comments etc.).

These records would be accessible to readers using a variety of interfaces, but including a search engine. This means that readers could search for information using a mixture of judgemental and content-based information. For instance: they could look for all papers on a subject which were judged to have very important results, so as to keep in touch with important developments; or they could look for papers with high originality however bad which mention a keyword if they needed some new ideas. In this way the reader can avoid being swamped by information with irrelevant characteristics and be assured of the quality of what is accessed without needlessly restricting the flow of information at an early a stage. The flow of quality information is not restricted in a generic way earlier in the publishing process, but instead the information allowing the later constraint of information is made available to readers so that they can constrain the information in the way that suits their needs.

At the moment readers are faced with a choice, either to read journals whose content will not be selected according to their precise needs or to select from the web or public archive on content, but have to wade through papers of low quality in ways that are important to the reader. The traditional system is illustrated in figure 1.

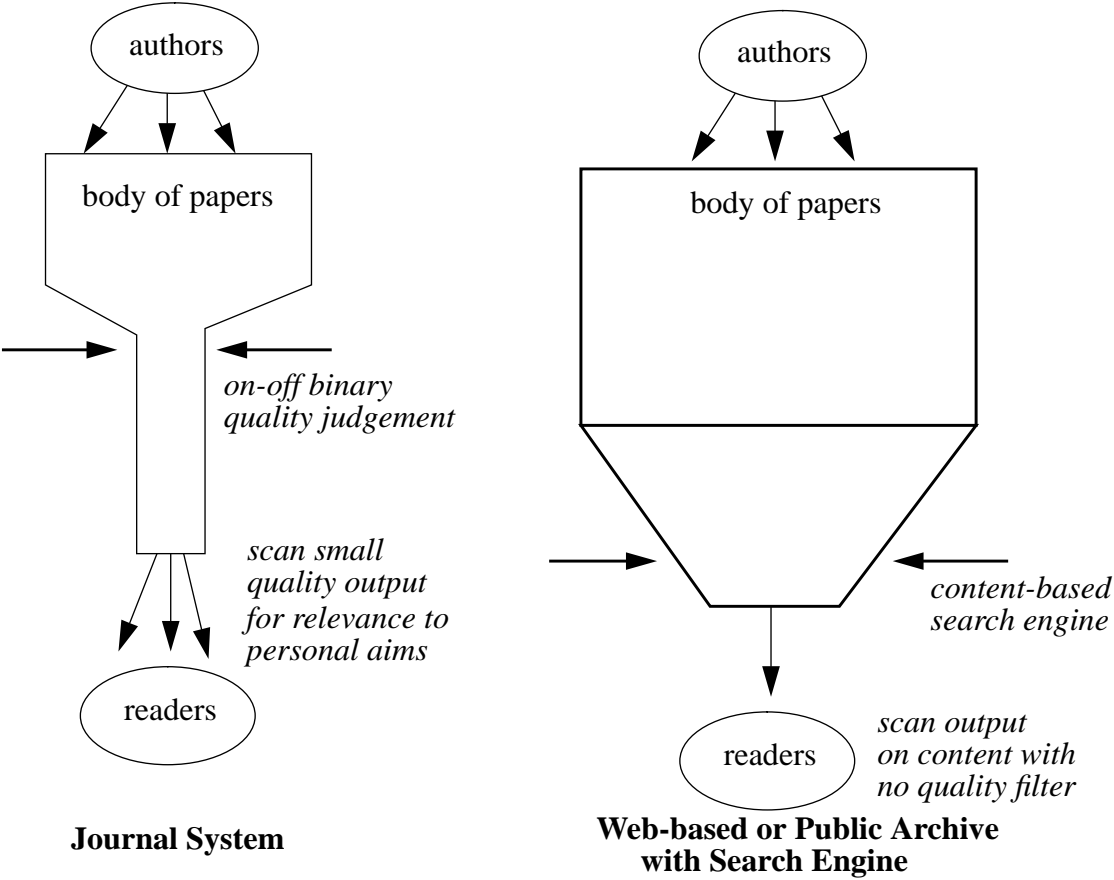

**Figure 1.** The existing journal and archive systems

Alternatively, the proposed system delays the selection process as long a possible and thus does not discard any information which may turn out to be of use later, but provides the user with the means to exclude the mass of papers in a way which meets the reader's needs at that time. The proposed system is illustrated in figure 2.

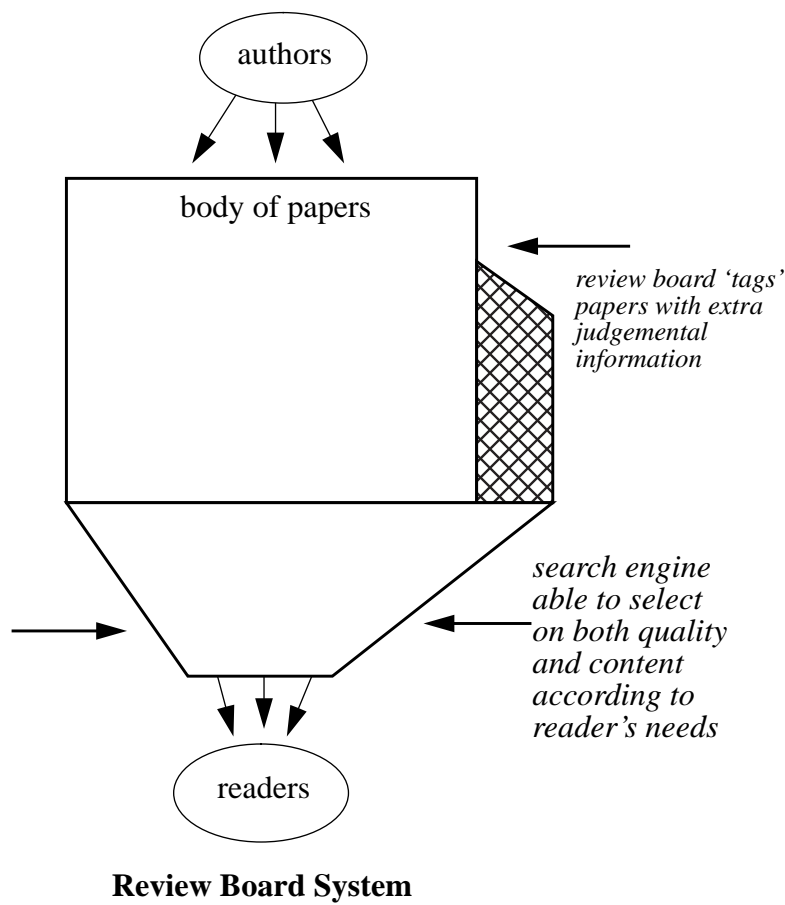

**Review Board System**

**Figure 2.** The proposed system

Of course the above illustrations are give a somewhat over-simple picture of the processes. There is a variety of journals, of different standards selecting on different purposes. Readers can use implicit and explicit knowledge of the quality and coverage of a journal to aid their search process. I presume that, if review boards were set up then there would be at least as great a variety of boards as there are presently of journals. Comparing like-with-like the journal system imposes premature, inflexible and unnecessary restrictions upon the flow of knowledge compared with a system of review boards.

The fact is that at the moment the only *practical and general* way to avoid scanning papers whose qualities are insufficient for one's purposes is to accept the selection of a journal whose criteria will not be identical to one's own. Of course, *if* one is in the happy position of having access to a manageable sized set of journals which (collectively) cover *all* ones needs then the journal system is sufficient, but many readers are not in such a position.

There is no reason why such a board should necessarily choose to restrict itself to those papers that are submitted to it by authors. It could equally well choose to review any papers that would be relevant to its readers (although if it were a paper published in a paper journal it would probably not be possible to supply an on-line link to it). Thus a review board could not only review new papers for its readers but also provide a new *view* of existing papers.

A system of review that is close to the above suggestions has bee made

## 4. The Potential Impact of Review Boards

The major advantages of this proposal are as follows:

1. It gives the readers access to richer judgemental information than merely whether it is deemed publishable or not in a particular journal. This information is already produced by reviewers it is just kept private.
2. Since this extra judgmental information can be provided in a form which can be utilised by search engines then the reader can set their own selection criteria using a mix of judgemental and content-based information based on their particular needs.
3. The fact that review boards would not own the papers means that they can give complementary judgemental views on some of the same papers. In this way they could help inform their readers about useful papers which would not normally be published in their field.
4. The fact that the presentation and ownership of papers is not the responsibility of the review board, and the fact that the review process is simplified meanss that review boards are less costly in academics time as compared to journals. Thus a greater variety of review boards will spring up in response to different needs in the academic community (and maybe beyond).
5. The system adds value to the totalityof academic knowledge in a way which is far more flexible than the existing system. For example once the judgemental information is published software writers can write their own new search engines to allow new ways of searching and relating knowledge.
6. The system is open, in that there is no private journal-author dialogue. Thus there would be less possibility of editors or reviewers coercing an author to change the content of a paper merely to suit their prejudices.

Possible disadvantages of the system could include:

1. That the time saved from the private review dialogue, the automation of the management of the review process, and the mark-up of text to fit a journal's style will be less than duplication of review of the same paper by different boards and at different times by the same board. I do no think that a new system of review boards will be swamped with unecessary work, for they will quickly develop rules to adjust their workload. Having said this I do think that in many cases the time spent re-reviewing a paper for a genuinely different audience and hence making it accessible to then, is time valuably spent.
2. The general standard of papers would drop as a result of the lack of a review dialogue. I do not think that a drop in standards would occur, merely that the mechanism of paper improvement would be different. The fact that a paper would be publicly judged with no chance of redrafting before the judgement is made public will mean that authors take greater care in their first drafts and get more feedback on the paper from colleagues, mailing lists and at workshops. I can envisage review services springing up that offer the author private feedback, but I think these would either be provided by academic institutions internally or charge for their services. If this occurs this would *relieve* the burden on the journals and review board.
3. That reliable sources of quality judgement may be lost. Review boards would gain status and permanence in much the same way as journals have. Initially they will be judged upon the eminence of the reviewers and institutions associated with it. Later they will also be judged by their output. I think it will take a long time before people switch allegiances from their trusted sources of quality information. If anything the danger is a transference to new review boards will be slower than is justified by the quality of their output. Also, if they are successful, I would expect the big names in journals to establish their own review boards under their known 'brand' to keep their readers and to try to retain the ability to charge using the resource of judgemental information they own as a result of their journal.

4. That academics in some fields will not have the skills to exploit the new system and so would be at a disadvantage. This is true to different extents in different fields. I doubt that academics in computer science will have much trouble with the new system, so that review boards will be set up in these sorts of disciplines first. The system would then 'percolate down' to other areas over time, just as has happened with web-journals.

Overall the ease with which review boards can be established and the flexibility of their nature will mean that the whole system will become more responsive and flexible. Some older, more entrenched institutions will be bypassed. However, in my experience most establishments have the ability to adapt if the need is suitably pressing. One assurance is that the whole system is adaptive and evolutionary: the new will only 'take-over' from the old if people vote for it 'with their feet'. The authors, reviewers and readers will all adapt their way of working so that they get the most out of the system, so that review boards will have to respond or suffer the fate of being ignored.

## 5. Pictures of Knowledge Development and Dissemination

In this section I relate the proposal to some different pictures of knowledge dissemination. The analogies that I draw do not prove that the proposal is better than the alternative but they are suggestive of this.

### 5.1. Evolutionary vs. Foundationalist

The traditional view of the development of knowledge (at least in the first half of this century in the west) was a foundationalist one. In this picture each piece of knowledge is validated and then relied on by subsequent pieces of knowledge in a cumulative fashion[1]. Such a system relies on there being one commonly applicable and acceptable set of validation criteria in a domain.

The archetype foundationalist subject is mathematics – there the criterion for validation of a piece of mathematics is clear: a piece of mathematics is valid and may be relied on if and only if the proof is correct, which is checkable (at least in theory). However, even in mathematics there is not clear agreement about what should be published because publication is largely dependent upon the importance of the results, and what is considered important is a subjective matter.

The evolutionary picture is one where academics are continually producing variations on older work and a selection process is acting upon these to give preference to the better ideas [3]. Those that are selected are more likely to be used and varied by future academics, so the *population* evolves in response to the selection pressures it is subjected to. These selection pressures are largely determined by the academics themselves (what lasts is what will appeal to academics over the years for good and bad reasons), but is ultimately grounded in the needs of the society that the academics inhabit.

---

1. e.g. as described in [10].

In the field of evolutionary computation (where artificial evolutionary processes are designed and studied), it is clear that there is some sort of trade-off between brittleness and cost. A greater and sharper selection pressure (as is applied in the breeding of show dogs) implies a quicker and cheaper convergence to a solution, but also means that one is more likely to get stuck in a sub-optimal solution. The reason for this seems to be that in hard problem spaces a good solution is sometimes reached by a variation on a very poor solution, so if one uses a 'crisp' selection criteria (i.e. only those above a certain threshold) one will select out this poor solution and never reach the good solution.

> *"…there is a necessarily a trade-off between concentrating the search in promising parts of the search space which increases the chance of finding local optima versus a wider ranging search which may therefore be unsuccessful but may find a more remote but better global optimum." [13] p.207*

In the field of the evolutionary population dynamics, there is a key theorem, 'Fisher's fundermantal theorem' [8] (updated and clarified in [15]), broadly it states that (under quite a broad range of assumptions) the rate of increase in the fitness of a population is proportional to the variation in that population. The proliferation of review boards and the increase in the use of paper archives will promote that variation.

Paper archives do go somewhat towards increasing the variation and have the effect of softening the selection process, but the proposal I describe goes further. Of course, the selection pressures are indispensable but perhaps a more graded selection process may promote the quality of knowledge by helping (in a small way) the evolutionary process.

### 5.2. Generic vs. Context-dependent Encapsulation of Knowledge

One big advantage of the journal system that would not be lost in a move towards review boards is in their context-specific nature. Each journal is able to employ reviewers, editors, review criteria editorial policy and style that is appropriate to the area it has chosen to cover and these can develop to keep pace with the developments in that area. Even the broad generic science journals such as *Nature* and *Science* have their known focus areas.

The big weakness of big paper archives and generic search engines is that they have to cover a very broad range of topics written about in a broad range of styles. Effective search strategies require the application of domain-specific meta-information. An analogy can be drawn with the 'No Free Lunch' theorems in computer science [5] which show (in a very abstract way and over an unnaturally large set of possible search spaces) that no search algorithm is better than others over all possible search spaces.

In the fields of AI and machine learning context-dependent learning and reasoning are increasingly seen to be powerful tools (e.g. [1]). Further there is increasing evidence that human learning and reasoning is inherently context-dependent (e.g. [6]). Review boards allow the introduction of context-dependency to a greater extent than journals because as well as covering different areas in different ways they can also cover the same papers in different ways for different audiences and purposes.

**5.3. Information vs. property**

The big advantage of information is that although good quality information is expensive to produce, once it has been produced there is no lower bound upon the cost of its replication. Intellectual property rights are precisely an attempt to levy a price on the replication of information[2]. This price can be explicit as in fees for access to journals and copyright fees or it can be implicit in that one may have to go to a certain place to get it.

Journals (whether paper or on-line) have tended to own their content, in that they control its appearance and own the copyright. The ownership of the copyright was necessary to protect the publisher and allow them to recoup their costs. Now that many of these costs have gone the review process does not have to be associated with the ownership of papers. The ownership of papers will necessarily impose (explicit or implicit) costs upon the distribution of knowledge. Review boards would discard this part of the journal tradition. For example it could become common that different boards review the same paper according to their own criteria developed to suit different audiences.

**5.4. Planning under uncertainty**

Another fruitful analogy can be drawn with results from AI planning systems, that are intended to work acceptably under conditions of uncertainty. One is faced with a large space of possible plans and some anticipated constraints imposed upon one one's choice imposed by the environment. The task is to choose a good plan. The algorithm chosen depends somewhat upon the extent of the uncertainty about the constraints. In a situation where the constraints are known with complete certainty then it is most efficient to apply the most restrictive constraint first to reduce the search cost. However if there is great uncertainty then it is sensible to apply the loosest constraint first and retain the flexibility about what would be the best plan to implement until the last possible moment[3]. In this way the delaying of constraints retains the maximum flexibility in order to be able to react to unpredictable changes in the environment.

With the advent of efficient search engines the search cost for readers is much reduced. Thus in fields where the journal will not be able to anticipate the selection criteria of its readers it is sensible not to prematurely select the content but to delay the selection process right up to the time the reader access the information. In this case it is better to 'tag' papers with judgmental information in a form that search engines can utilise and allow readers to choose the selection criteria themselves rather than attempt to do the selection on behalf of a disparate set of readers prematurely. Previously the cost of implementing such a delayed search was prohibitive and this is why the journal system was appropriate, now with the shift in the cost structure this delayed selection is not only possible by fairly easy to implement.

Of course, if a journal is in the happy position of being able to anticipate its readers selection criteria with a high degree of faithfulness, then the (now minimal) search cost they save its readers may justify its existence. I think that for many journals this is not be the case, and that in the future the search cost saved will diminish as compared to the flexibility of review.

---

2. As well as control its use and development.
3. This is expressed in the literature in terms of 'contigency planning' as in [4] – the greater the uncertainty the more contigency planning is required.

## 6. Practical Steps for the Establishment of Electronic Review Boards

An illustration of a possible review process is illustrated below in figure 3. In the subsections below I briefly outline what is necessary to implement it.

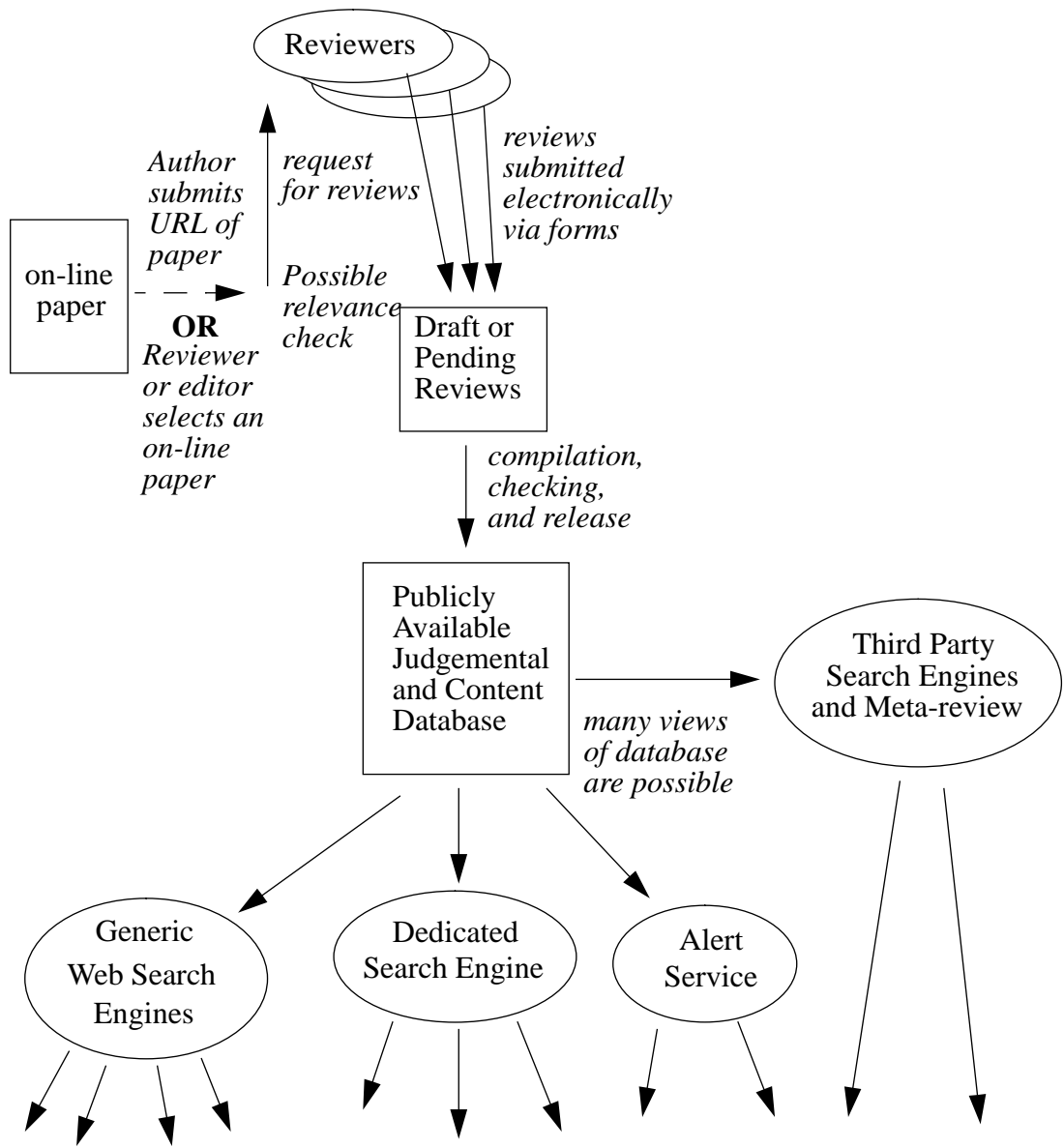

**Figure 3.** An illustration of a possible review board process

As is frequently the case, the organisational issues concerning the working habits of people dominate the merely technical difficulties in setting up review boards as described herein. So I will deal with the technical requirement first.

## 6.1. Software required

### 6.1.1 Format

The format of the judgemental information that is to be made available to the public will constrain the uses to which it can be put. For this reason I suggest that the review information be stored as viewable web pages in HTML (or its successor XML), with any machine searchable fields be included (or duplicated) in the header section of the HTML using meta tags, for example: <META NAME="author-name" CONTENT="Edmonds, Bruce">. In this way many search engines can be easily adapted to index and search on the information (e.g. Harvest), as well as being browsed by the public and indexed by public search engines in the normal way.

The fields should conform to an accepted standard like ReDIF [14] and should include: standard information such as author, title, institutions, publication date, keywords and URL; information on the review board such as title, URL, classification codes, keywords, and e-mail of maintainer; and the judgemental information which could be anything the board desires, including average originality rating, number of reviewers, review date, average importance rating, standard of presentation, soundness and reviewer comments. Some of these, such as the title, can only be searched using key words and phrases but other data, such as level of presentation, are ordinal in nature and should be represented numerically so that the searcher can specify a minimum level (e.g. presentation > 2).

If there was a core of these fields that were agreed upon by a number of such boards this would ease the learning process by readers and enable secondary indexing and search engines to be developed.

### 6.1.2 Input scripts

To ease the creation of the database of judgemental information, scripts should be written that allow reviewers to enter their review and translate it into the required format (merging any numerical judgements with any existing ones). This could also preform some basic checking on the information (like the presence of the paper at the specified URL). The information update could then be released into the boards database with minimal extra checking required. A security system should allow only accredited reviewers to input reviews and an internal audit trail is needed of which reviewers submitted what in order that the board may trace actions done. The facilities for these scripts and associated security measures are widely available.

### 6.1.3 Query server

A search engine will need to be installed (either by the review board or a third party) to deliver the information in an accessible way to the reader. Ideally this should be customized to allow the easiest access to the judgemental data. The reader would then merely follow the suggested links to access the papers they chose. Given the public accessibility of the database information a variety of different interfaces could be provided for the reader. A suitable indexing and serch system is harvest [9].

### 6.1.4 Required public web pages

In addition to the database of judgemental database and the search engine with its interface, there would need to be a few web pages to provide the context for the service. These would include: a title page, a page specifying the reviewers and supporting institutions, and a page describing the meaning of the judgemental information provided (which would typically involve some description of the process whereby it was derived) and some help pages.

The actual pages containing the information should be publicly linked to the title page (via an index page) so that the public and public search engines can reach them, in addition to access through the provided search engine.

### 6.1.5 Alert service

An additional service that would be useful to readers would be a flexible 'alert' service that would e-mail subscribers whenever new review information was released that met their selection criteria. This would have to perform a regular search for each subscriber according to the stored selection criteria on any new or updated pages provided and then email the corresponding subscribers with a notification and location of the page. Some of the on-line indexing services already provide such a service (e.g. uncover [16]), but I have not seen publicly available software that would do this.

### 6.1.6 Internal mailing list or discussion forum

In order to develop and preserve and indentifiable coherency of approach the members of the review board need to be in constant touch with each other. This may be facilitated by a private mailing list or discussion forum. Notification of incoming reviews may be automatically made available to board members and discussions about editorial policy may be done using these means. Software for the management of mailing lists are freely available (e.g. majordomo [13] and hypermail [11])

## 6.2. Personnel

Given that the most of the above software is already in existence and merely need to be put together (with the exception of the additional alert service), the biggest barrier to the establishment of a review board is the initial gathering of people to make it happen.

### 6.2.1 Review board

First and foremost, one would need a body of academics to do the reviewing. These would need to be sufficient in number, dedication and coherence to produce a sufficient volume of quality review information so as to be useful to its readers. It is likely that these reviewers would have to spend some time creating reviews before the service was announced. It is likely (but dependent on their own policy) that two or three reviews would have to be collected on each paper before the judgement were publicly released.

### 6.2.2 Managing editor(s)

Due to the one-shot nature of the review process, the job of managing the review process is greatly lessened and far more amenable to partial automation. However (dependent on the board's exact policy) there is still a residual job in checking the information before it is released, leading conversations on policy and direction with the reviewers and occasionally requesting reviews on papers that need more for the information to become public.

### 6.2.3 Software/archive maintenance

Lastly someone has to manage the software half of the system, take backups of the information and generally maintain it. Such a job would presently involve a computer specialist, but in the future is more suited to a person whose expertise is in the management of archival information: a librarian.

### 6.3. Establishment of status

Once a board is established, the reader's choice to use the service will be largely based upon their own experience of it. A review board has the same problems as new journals, namely credibility. I would anticipate that many of the traditional solutions would be used: known and trusted reviewers, prestigious institutional associations, and basic advertising.

Of course, ultimately there could be a Review board of Review boards to help readers choose!

## 7. Conclusion

The shift in the cost structure of publishing caused by cheap computing power and the internet means that a more flexible systems of information filtering can now be implemented, allowing the readers to set selection criteria that meet their *individual* needs. This can be done by publishing some of the judgemental information that reviewers *already* produce, in a form that readers can utilise using a search engine. This *delayed* application of selection ensures that the system is as flexible as possible in helping the readers find the information they want.

The separation of the review process from that of paper ownership and the simplification of the review process from a closed dialogue to a simple one-shot evaluation, enables the semi-automation of the management and hence further reduces costs. This should enable a large variety of review boards to be set up, segmented not only by subject matter but also by approach and information offered to the reader.

The final balance of time saved and wasted using a system of review boards compared to the journal system will only be discovered in practice. This I intend to do in the *near* future. However it is indicative that the main work left is that essential to the academic process: writing, selection and reading.